# Dose Delivery Verification

*S. Safai*
Paul Scherrer Institut, Villigen, Switzerland

**Abstract**
This paper focuses on some dosimetry aspects of proton therapy and pencil beam scanning based on the experience accumulated at Paul Scherrer Institute (PSI). The basic formalism for absolute dosimetry in proton therapy is outlined and the two main techniques and equipment to perform the primary beam monitor chamber calibration are presented. Depth–dose curve and lateral beam width measurements are exposed and discussed in detail, with particular attention to the size of the ionization chamber and the characteristic of scintillating–CCD dosimetry systems, respectively. It is also explained how the angular–spatial distribution of individual pencil beams can be determined in practice. The equipment and the techniques for performing regular machine-specific quality checks are focused on (i) output constancy checks, (ii) pencil beam position and size checks and (iii) beam energy checks. Finally, patient-specific verification is addressed.

## 1 Introduction

Dosimetry plays an essential role in the clinical activities of any centre that offers radiation therapy as a modality to treat patients afflicted by tumours. After the installation and tuning of any treatment unit, dosimetry is necessary first to accept and then to characterize such a unit in what are typically referred to as *acceptance* and *clinical commissioning*, respectively. After a successful conclusion of the commissioning phase, dosimetry is then required on a regular basis as part of the quality assurance programme. As part of that programme, dosimetric quality and consistency checks are typically repeated on a daily, weekly, monthly and yearly basis to provide the confidence that the system is behaving as expected. Particle-therapy centres are not an exception and adhere strictly to this well-established practice.

In our overview on dosimetry in particle therapy we will follow the sequence outlined above by first looking into the dosimetry equipment and techniques for clinical commissioning (Sections 2 and 3) and then into those for periodic checks (Section 4) with particular attention to those aspects relevant to proton therapy and pencil beam scanning (PBS).

## 2 Absolute dosimetry

The most relevant task of absolute dosimetry at commissioning is the calibration of the primary beam monitor chamber (BMC). The primary beam monitor chamber—usually a large parallel-plate ionization chamber situated in the nozzle of a gantry—is typically calibrated in terms of monitor units (MUs) per dose–area product (MU/$D_w A$) or alternatively in terms of MUs per proton (MU/p) [1]. The preferred choice is, to a large extent, dictated by the requirements of the treatment planning system (TPS), which, to produce a desired dose distribution, could predict either the dose per pencil beam (most commercially available TPSs) or the number of protons per pencil beam (e.g. the in-house TPS PSIPLAN of PSI). The former would require a MU/$D_w A$ calibration that could be derived with ionization chamber measurements based on IAEA TRS-398 [2], the latter a MU/p calibration that could be derived with Faraday cup measurements. In what follows we will first review the basics of the TRS-398 Code of Practice for protons and then explain the two major techniques to calibrate the BMC.

## 2.1 Code of practice: the basic formalism

ICRU 78 [3] adopted the IAEA TRS-398 Code of Practice for reference dosimetry in proton therapy. The reader is therefore referred to the latter for a comprehensive overview of reference dosimetry in particle therapy. What follows is a short overview from that report.

The basic formalism according to TRS-398 described the absorbed dose to water $D_{w,Q}$ for a beam quality $Q$ in the following way:

$$D_{\mathrm{w},Q} = M_Q N_{D,\mathrm{w},Q_0} k_{Q,Q_0}. \tag{1}$$

Here $M_Q$ is the instrument reading at user beam quality $Q$, corrected for all influence quantities, $N_{D,\mathrm{w},Q_0}$ is the absorbed dose to water calibration coefficient for calibration beam quality $Q_0$ (usually $^{60}$Co) and $k_{Q,Q_0}$ is the beam quality factor to correct for effects of differences between calibration beam quality $Q_0$ and user beam quality $Q$.

As primary standard laboratories have no, or very limited, access to proton beams, the reference beam quality remains $^{60}$Co and the values for $k_{Q,Q_0}$ tabulated in TRS-398 for protons are derived by calculation rather than experimentally, which introduces additional uncertainties in reference dosimetry with protons.

Typically reference dosimetry in a proton beam is performed by measuring the dose with cylindrical ionization chambers placed in the plateau region of a spread-out Bragg peak (SOBP). The beam quality index for the proton beam is defined as the residual range $R_{\mathrm{res}}$ in g/cm² at a measurements depth $z$, with

$$R_{\mathrm{res}} = R_{\mathrm{p}} - z. \tag{2}$$

Here $R_{\mathrm{p}}$ is the practical range, i.e. the depth at which the absorbed dose beyond the Bragg peak falls to 10% of its maximum value.

## 2.2 Beam monitor chamber calibration

### 2.2.1 With ionization chambers

To calibrate the BMC, the use of SOPBs, even though it is the standard for reference dosimetry, may not be advisable for proton pencil beam scanning under certain conditions, as first pointed out by Jäkel *et al.* [4] for heavy ions. In fact, its applicability depends on the location of the energy modulation system along the beam line with respect to the BMC, as explained below.

(a) If the energy modulation is performed downstream from the BMC, for instance with the use of range shifters in the nozzle, then the BMC will always 'see' the same energy for all the energy layers delivered in a given SOBP. As such, only one calibration value has to be determined for a given energy tune. In this case BMC calibration with reference dosimetry in the middle of SOBPs is applicable and should be performed for every energy tune.

(b) If, on the other hand, the energy modulation is performed upstream from the BMC, for instance with the use of a degrader just after the accelerator, then the BMC will 'see' different energies, a different one for every energy layer in a given SOBP. Since, in this case, each energy layer corresponds to a different energy tune, the calibration has to be performed individually for every deliverable energy or at least for a subset of deliverable energies.

Hence, the BMC calibration with reference dosimetry in the middle of SOBPs is not advisable in this case. A much more practical approach is the calibration performed with use of small parallel-plate ionization chambers (e.g. Markus chambers) placed at shallow depth in water, e.g. at $z$ equal to 2 g/cm$^2$, in the middle of a 10 × 10 cm$^2$ monoenergetic energy layer, to be repeated for all, or a subset of, deliverable energies. Equation (2) is then used to compute the beam quality index for a given energy layer in order to extract the corresponding $k_{Q,Q_0}$. With this method parallel-plate ionization chambers are preferred to cylindrical chambers since the measurements are performed in a gradient region. As such, particular attention should be given in positioning the chambers at the desired depth $z$.

### 2.2.2 With Faraday cups

The use of Faraday cups is the direct way to calibrate the BMC in terms of MU/p.

Typically the Faraday cup is placed at the exit of the nozzle in air. All energy tunes, or a subset of deliverable energy tunes, are individually delivered. After delivery, the number of protons is determined directly by the measured charge in the Faraday cup; this number is recorded together with the number of MUs registered by the BMC for a given energy. The ratio between the two numbers will then provide the calibration.

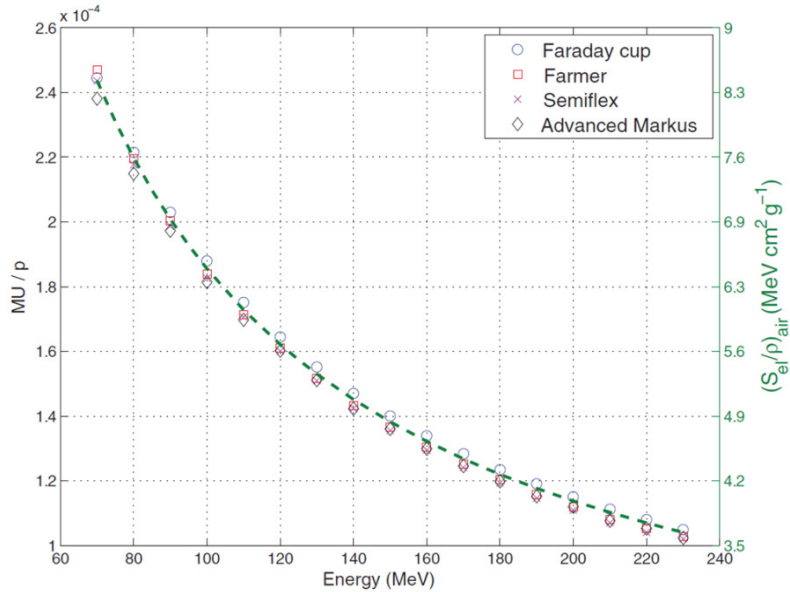

**Fig. 1**: Comparison between a BMC calibration performed with a Faraday cup and the calibration performed with ionization chambers [1].

At Paul Scherrer Institut (PSI) we compared the calibration performed with a Faraday cup with the calibration derived with different ionization chambers (Fig. 1). The calibrations agree within 3%. Details of this comparison have been published by Gomà *et al.* [1].

### 2.2.3 Validation

Regardless of the method used to calibrate the BMC, reference dosimetry in SOBPs following TRS-398 remains a valuable way to validate the calibration determined with that particular method and to introduce additional correction factors if a significant discrepancy is observed. As an example, at PSI we opted for the Faraday cup method in Gantry 2, which has an upstream energy modulation system as described in Section 2.2.1(b). After the BMC calibration with the Faraday cup, the dose was verified following TRS-398 in the middle of several SOPBs located at different depths. The difference between the measured dose and the expected one was then used to introduce an energy-dependent correction factor in the calibration.

## 3 Relative dosimetry

At clinical commissioning *relative* dosimetry plays an essential role in the collection of the reference dosimetric beam data required by the TPS for dose calculations. The type and quantity of the data to be collected depend on the requirement of a specific TPS. In general, a comprehensive set of pencil beam integral depth–dose curves and lateral beam widths (i.e. spot sizes) is needed, typically measured for a subset of the deliverable energies. We will illustrate the type of equipment and techniques to perform such measurements.

### 3.1 Depth–dose curve measurements

Integral depth–dose curves for monoenergetic pencil beams are usually measured with large circular parallel-plate ionization chambers with a diameter ($\varnothing$) of at least 8 cm. These chambers have an excellent resolution in depth and are large enough to collect the dose deposited by both primary and secondary particles. The chamber is placed in a water tank and while a pencil beam is delivered, either with constant beam ON or on a spot-by-spot basis, the chamber is moved along the beam axis to measure the entire Bragg peak curve (Fig. 2). At each position in depth the integral signal measured by the chamber is recorded and later normalized with either the MU measured by the BMC or the signal from a large reference chamber positioned at the entrance of the tank.

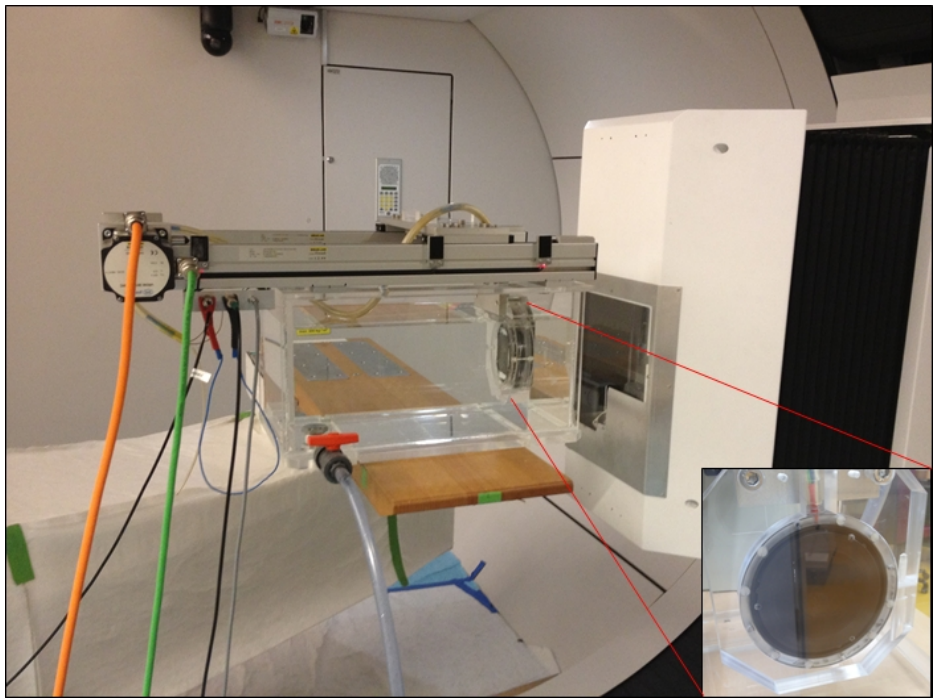

**Fig. 2:** Experimental set-up at PSI Gantry 2 for range measurements in water

A question that is often asked is what should be the ideal size of the chamber for such measurements. Figure 3 shows in a logarithmic scale the lateral profile of a pencil beam measured at mid range in water for a 150 MeV pencil beam. The primary protons contribute mainly to the central envelope of the distribution and can be well described with a Gaussian function (the first Gaussian in the plot). The sigma of this function is often used to characterize the lateral pencil beam width (also referred to as 'spot size'). The secondary particles, as a result of nuclear interactions of the primary beam in the medium, deposit dose not only in the central envelope but also outside that region and create a so-called *halo* of deposited dose around the primary beam. The halo is mainly produced by large scattered secondary protons. In its most simplistic form the halo can also be described with a Gaussian function but with a significantly larger sigma [5]. The sigma of this second Gaussian could be up to 2-cm large [5]. Figure 3 shows that the second Gaussian can well describe the beam halo up to 4 cm from the central axis. Beyond that point there could be additional dose deposited that would need a better mathematical description. From this we learn that an 8-cm (∅) circular ionization chamber could miss some of the dose deposited at larger radius and that a wider chamber could be advisable. On the other hand, if a treatment planning system is unable to properly describe the additional dose deposited outside that boundary, e.g. if only a two-Gaussian model is implemented, then an 8-cm chamber could be sufficient and the use of a larger one could be counterproductive. Hence, the answer to the original question is not so straightforward and depends also on how well the TPS is able to describe the halo. At PSI we decided to use an 8-cm chamber. Figure 4 shows the comparison between 8-cm and a 12-cm chambers for five different energies measured at PSI. A small but significant difference is observed only at mid range for high-energy beams.

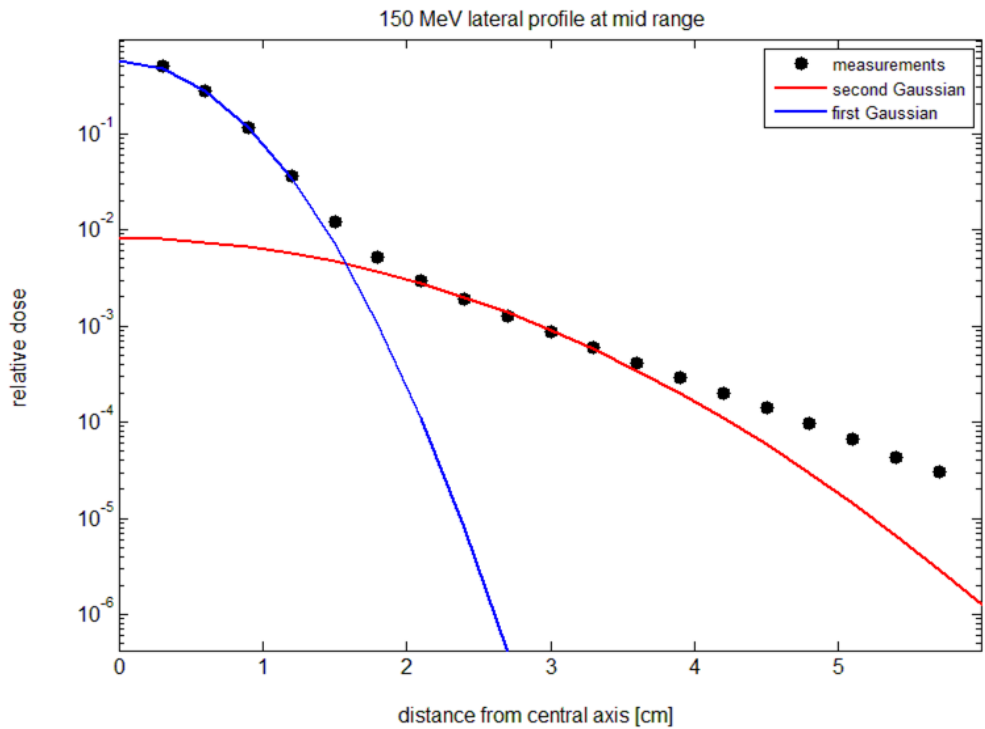

**Fig. 3:** Transversal profile of a 150 MeV pencil beam measured in water at mid range

## 3.2  Lateral beam width measurements

In this section first a brief exposure of the fundamentals of pencil beam propagation in air is given followed by how to determine in practice the spot size and angular–spatial distribution of individual pencil beams.

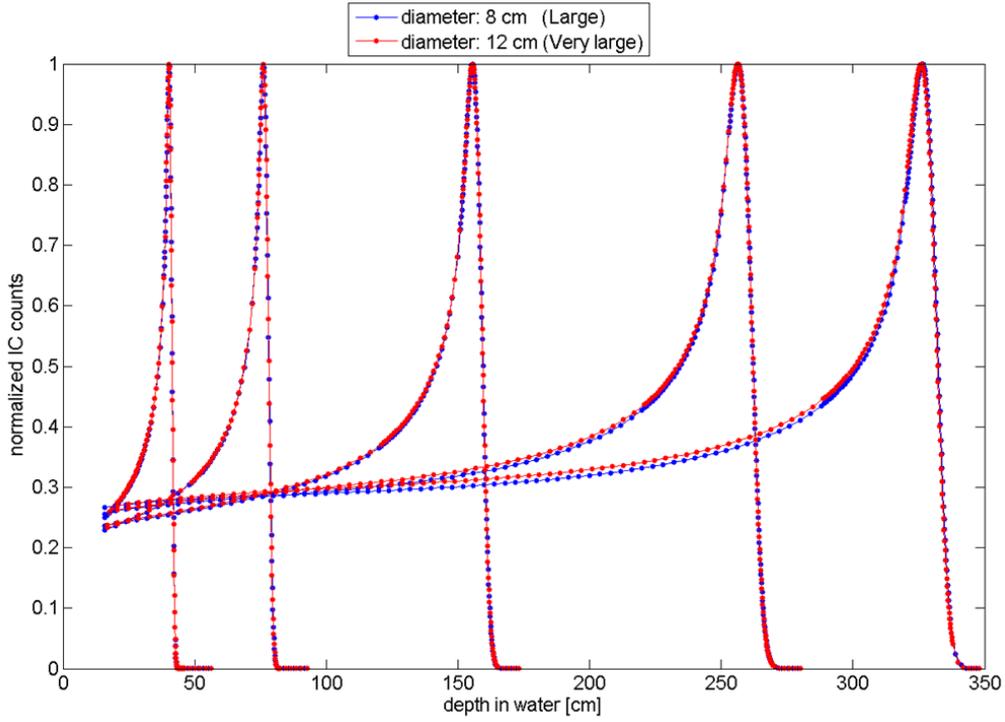

**Fig. 4:** Depth–dose curves measured in water with 8-cm and 12-cm-large parallel-plate ionization chambers

### 3.2.1 Pencil beam propagation in air: generalized Fermi–Eyges theory

Generally the beam optic of a PBS gantry is designed and tuned to bring the waist of the beam as close as possible to the gantry isocentre, i.e. the spot is smallest at isocentre. In reality multiple Coulomb scattering (MCS) in the nozzle and air moves the waist further upstream from the isocentre, in particular for low-energy beams, for which MCS is more pronounced. As the lateral size of an individual pencil beam changes as the beam propagates in air, a unique value to describe the size of such a beam is, in general, not sufficient. Generalized Fermi–Eyges theory can well describe the propagation of a pencil beam in air by parametrizing the angular–spatial distribution of the pencil beam with three parameters, the so-called *moments* of the distribution, $A_0$, $A_1$ and $A_2$ [6, 7]. The angular–spatial distribution, ASD, represents the Eyges solution to Fermi's diffusion equation and can be described as follows (for the $x$ coordinate):

$$\mathrm{ASD}_x(x,\theta,z) = \frac{1}{\pi\sqrt{A_0 A_2 - A_1^2}} e^{-\frac{(A_0 x^2 - 2A_1 x\theta + A_2 \theta^2)}{(A_0 A_2 - A_1^2)}}. \tag{3}$$

Here $\mathrm{ASD}_x$ describes the distribution of the protons within a spot both in angle and in position projected onto the $x$–$z$ plane at a given longitudinal position $z$. In general, $A_0$, $A_1$ and $A_2$ are a function of $z$ and they represent the doubled angular variance, doubled covariance and doubled spatial variance, respectively. Hence, we can write

$$A_0(z) = 2\sigma_\theta^2(z), \tag{4}$$

$$A_1(z) = 2\mathrm{Cov}(x,\theta,z), \tag{5}$$

$$A_2(z) = 2\sigma_x^2(z), \tag{6}$$

where $\sigma_x$ and $\sigma_\theta$ is the spatial and angular spreads, respectively. As in general the multiple Coulomb scattering in air is small, the propagation of the beam at the exit of the nozzle and in the proximity of the isocentre can be expressed as it would propagate in vacuum, i.e. with

$$A_2(z) = A_{0,0}z^2 + 2A_{1,0}z + A_{2,0}, \qquad (7)$$

where $A_{0,0}$, $A_{1,0}$ and $A_{2,0}$ represent the initial moments of the distribution at the reference position $z = 0$. When $A_2(z)$ is known, then the spot size (expressed with $\sigma_x$) at a given position $z$ can be computed with Eq. (6), i.e.

$$\sigma_x = \sqrt{A_2(z)/2}. \qquad (8)$$

### 3.2.2 Angular–spatial distribution determination in practice

Nowadays treatment planning systems can describe the beam propagation in air with Eq. (7) or with similar equations. From Eq. (7), we learn that the doubled spatial variance $A_2$ is a quadratic function of $z$ and that the initial moments are the coefficients of this function.

Experimentally the coefficients are determined by measuring $A_2$ at, at least, three different planes in $z$. A quadratic fit to the measured $A_2$ as a function of $z$ will provide the coefficients, i.e. the initial moments of the angular distribution, which will be used by the TPS to predict the spot size at any other plane in air via Eqs. (7) and (8) (see Fig. 5). The values of $A_2$ for the fit are typically determined experimentally by measuring high-resolution 2D lateral profiles of individual pencil beams at those planes. A 2D Gaussian fit is then performed on the 2D lateral profiles to obtain $\sigma_x$ and $\sigma_y$. Equation (6) and the analogous equation for $y$ are then used to compute $A_2$ at those planes.

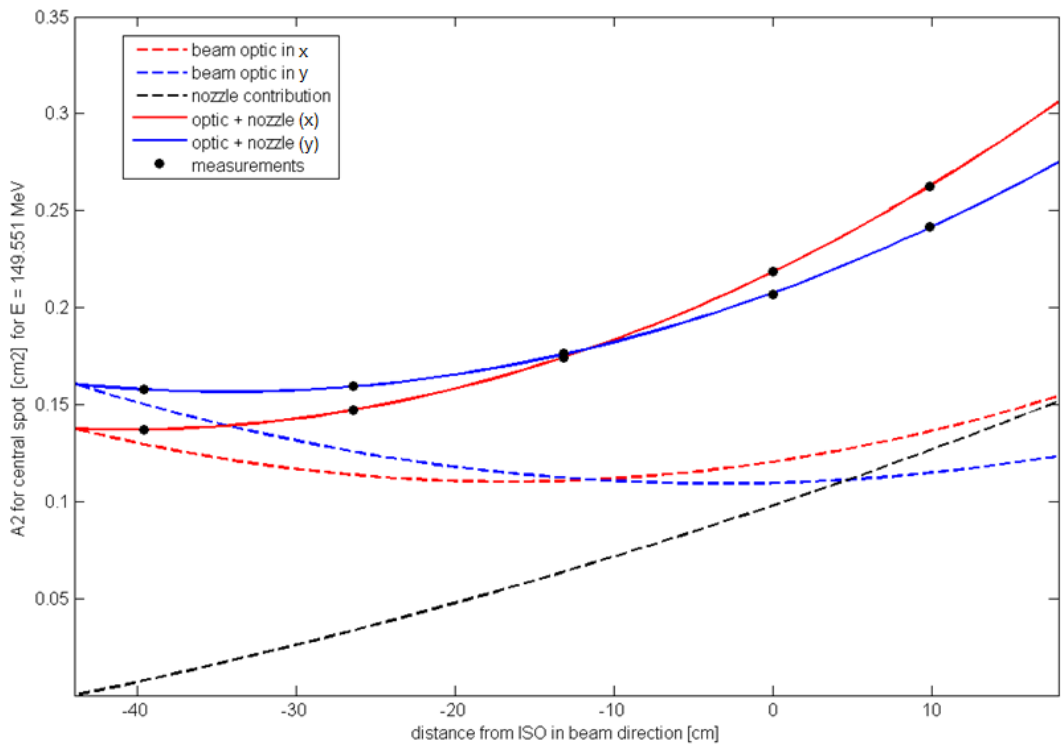

**Fig. 5:** Double spatial variance $A_2$ for a 150 MeV pencil beam measured in air at PSI Gantry 2 (dots), and the quadratic fit through the solid lines. The black dashed line shows the nozzle contribution due to the MCS of the nozzle alone. The dashed coloured lines are derived from the difference between the fit and the nozzle contribution and can be interpreted as the propagation of $A_2$ due to the beam optic alone.

For this kind of measurement it is advisable to use high-resolution detectors, such as scintillating screens or Gafchromic films (e.g. EBT3). Scintillating screens are usually used in combination with a mirror and a CCD; therefore hereafter we will refer to this equipment as *scintillating–CCD dosimetry systems* (Fig. 6).

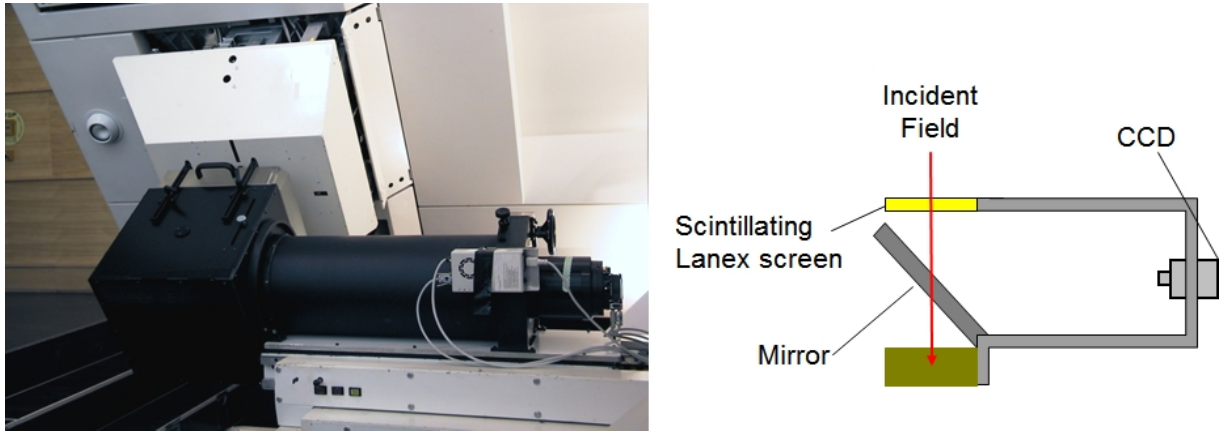

**Fig. 6:** Scintillating–CCD dosimetry system at PSI Gantry 1 (left) and a schematic representation (right)

It should be noted that the response of both Gafchromic films and scintillating screens depends on the energy and the linear energy transfer (LET – the energy transferred to the medium per unit of path length) [8, 9]. The response decreases with decreasing energy and increasing LET. Therefore, a pencil beam depth– dose curve measured with such systems shows a supressed Bragg peak compared to a curve measured with ionization chambers. This phenomenon is documented in the literature and has been often called the *quenching effect*. The quenching effect is irrelevant for transversal dose distribution measurements when the energy spectrum across the measurement plane can be assumed constant, which is generally the case for spot-size determination in air [10].

Scintillating–CCD systems have the advantage compared to Gafchromic films to be linear with dose and to have an electronic readout providing faster and more accurate results. With these systems a large quantity of data can be recorded and analysed almost simultaneously, representing one of the most efficient ways of collecting the necessary dosimetric data at commissioning.

## 4 Periodic checks

### 4.1 Machine-specific dosimetry

For PBS there are three main important dosimetry aspects that have to be verified constantly, on a daily, monthly and yearly basis, i.e.:

   i.   absolute dose (output constancy check);
   ii.  pencil beam position and size (including alignment at isocentre);
   iii. beam energy (range measurements).

The rationale for these checks is to identify as early as possible problems with: (i) the monitor calibration and/or in general with the system, (ii) the scanning system and/or beam line optic and (iii) the energy selection system and/or beam line.

### *4.1.1 Output constancy checks*

Output constancy checks are typically performed with reference ionization chambers, either in a phantom (daily checks) or in water (yearly checks), in the middle of flat dose distributions (i.e. SOBPs).

### *4.1.2 Pencil beam position and size checks*

Scintillating–CCD systems, strip chambers, Gafchromic films and amorphous-Si detectors are among the devices that can be used to perform constancy checks of the beam size and of the accuracy of beam positioning. Compact systems like strip chambers could be used on a daily basis, while larger systems like the scintillating–CCD systems could be employed on a monthly or yearly basis, even though it is difficult to generalize. As a matter of fact, there are now new commercial products coming to the market designed to use scintillating–CCD systems for daily quality assurance, which are quite promising.

The performance of the scanning system, i.e. the accuracy of pencil beam positioning, is probably the one more prone to changes over time than any other system (e.g. energy-selection system); at least, this is our experience at PSI. The accuracy of beam positioning is typically gantry angle and energy dependent and could fluctuate from day to day. It could also depend on the ramping scheme of the beam line and the sequence in which the energy layers are delivered. Therefore, particular attention should be given to the frequency and comprehensiveness of the consistency checks for this system. Both the alignment of the beam at isocentre should be verified (absolute beam position) as well as the relative position between neighbouring spots. The homogeneity of large energy layers is quite sensitive to the precision of the placement of the individual pencil beams within the layers. Hence, the delivery of such layers in air, covering the lateral extent of the scanning region, on a 2D high-resolution detector (e.g. scintillating–CCD system), is an effective way to verify the performance of the scanning system.

### *4.1.3 Beam energy checks*

The energy is indirectly checked by verifying the range and/or the shape of the Bragg curve. Range measurements in water as described in Section 3.1 are time consuming and are usually repeated only on a yearly basis. Multilayer ionization chambers (MLICs), on the other hand, can record a full Bragg curve extremely fast in a single measurement [11]. If, on top of that, the MLIC can be synchronized with the beam delivery then hundreds of energies could be measured in a few minutes. MLICs are cross-calibrated against measurements in water and the reproducibility of the measured range is very high, well below the typical tolerance of 1 mm for range checks. If well integrated, a MLIC could be a powerful tool to perform daily range measurements or it can be employed for comprehensive weekly or monthly checks.

The use of wedges in combination with 2D detectors (e.g. scintillating–CCD systems) is another alternative to perform energy checks, even though the number of energies that can be verified in the same session is small. Another method is to look at the ratio of the signal measured by at least two small detectors (ionization chambers or diodes) placed at different depths. The ratio is than compared to a reference value, which is characteristic for a given energy.

### 4.2 Patient-specific dosimetry

Patient-specific verifications are typically performed by measuring 2D dose profiles at different planes with 2D arrays of ionization chambers in phantoms. For comparison the planned dose is recalculated for that particular phantom and geometry used during verification. Sometimes 2D arrays are only used to verify the relative dose distribution. In this case, the absolute dose is additionally verified with calibrated ionization chambers (e.g. pin-point chambers) in a water phantom for one of a few selected reference points. The field under consideration would have to be applied at least twice, one for relative and one for absolute dose verification. Ideally the dose should be verified under the proper treatment gantry angle for that field, but this would require a rotatable phantom. When a rotatable phantom is not available, then, on top of the field verification under a reference angle, it is advisable to run the field also under the treatment angle but without performing a dose measurement, just to verify that the field is deliverable. This is for the reasons pointed out in Section 4.1.2; the dedicated monitor in the nozzle could identify potential problems of spot positioning for that field.

Each individual field is typically verified and the output corrected if a discrepancy is observed between measured and planned doses. At PSI, only after several years of experience and improvements in the in-house dose calculation engine did we reach the confidence to drop the verification of every individual field planned for Gantry 1. On the other hand, Gantry 2 has been in operation only for a few years and, therefore, as of now, all planned fields for this gantry are being verified.

### *4.2.1 Log file analysis*

Dose calculations based on the parameters registered in log files during beam delivery, such as spot position and delivered MUs, could become a relevant tool in the verification of the delivered dose and could reduce the amount of measurements to be performed for each planned field prior to the treatment [12]. This approach could also verify on a daily basis the delivered dose and could therefore be used to adapt the treatment if necessary.